\documentclass{aa} 
\usepackage{graphicx, amssymb,epsfig} 
\usepackage{times}
\usepackage{txfonts}
\usepackage{multirow}
\usepackage{natbib}
\usepackage{longtable}
\usepackage{color}
\usepackage[usenames,dvipsnames]{xcolor}
\usepackage{ulem}
\newcommand\T{\rule{0pt}{2.6ex}}
\newcommand\B{\rule[-1.5ex]{0pt}{0pt}}

\begin{document} 
\title{Dust evolution in the transition towards the denser ISM:\\
impact on dust temperature, opacity, and spectral index}
\author{M. K{\"o}hler\inst{1}\and
              N. Ysard\inst{1}\and
              A. P. Jones\inst{1}}  

  \institute{Institut d'Astrophysique Spatiale (IAS), Universit\'e Paris Sud \& CNRS, B\^at. 121, 
       Orsay 91405, France  } 
 
\date{Received ..; accepted ..} 

\abstract
  {Variations in the observed dust emission and extinction indicate a systematic evolution of grain properties in the transition from the diffuse interstellar medium (ISM) to denser molecular clouds.}
  {The differences in the dust spectral energy distribution (SED) observed from the diffuse ISM to denser regions, namely an increase in the spectral index at long wavelengths, an increase in the FIR opacity, and a decrease in temperature, are usually assumed to be the result of changes in dust properties. We investigate if evolutionary processes, such as coagulation and accretion, are able to change the dust properties of grains in a way that is consistent with observations.}
   {We use a core-mantle grain model to describe diffuse ISM-type grains, and using DDA we calculate how the accretion of mantles and coagulation into aggregates vary the grain optical properties. We calculate the dust SED and extinction using DustEM and the radiative transfer code CRT.}
   {We show that the accretion of an aliphatic carbon mantle on diffuse ISM-type dust leads to an increase in the FIR opacity by a factor of about 2 and in the FIR/submm spectral index from 1.5 to 1.8, and to a decrease in the temperature by about 2 K. We also show that the coagulation of these grains into aggregates further decreases the temperature by 3 K and increases the spectral index up to a value of $\sim$2. The FIR opacity is increased by a factor of 3 (7) for these aggregates (with an additional ice-mantle) compared to the diffuse ISM-dust.} 
   {Dust evolution in the ISM resulting from coagulation and accretion, leads to significant changes in the optical properties of the grains that can explain the observed variations in the dust SED in the transition from the diffuse ISM to denser regions.}

\authorrunning{K{\"o}hler et al.}
\titlerunning{}

\maketitle


\section{Introduction} 
\label{intro}

Indications of dust evolution from the diffuse interstellar medium (ISM) to molecular cloud environments ($n>10^3$ ${\rm H/cm^{3}}$) have recently been observed with the {\it Herschel} \citep{pilbratt-et-al-2010} and {\it Planck} \citep{lamarre-et-al-2010} space observatories in the far infrared (FIR) and submillimetre (submm) wavelength range. 
Earlier, dust evolution was already indicated with ISO, Spitzer, SPM/PRONAOS, and IRAS data.
For example, a temperature decrease of large grains, which are in thermal equilibrium with the radiation field, from 20.3~K in the diffuse ISM \citep{planck-XI-2013} to less than 14~K in denser regions \citep[e.g.,][]{lagache-et-al-1998,stepnik-et-al-2003,abergel-et-al-2011} was observed together with an increase in the dust opacity by a factor of 2 to 4 in the FIR/submm \citep[e.g.,][]{stepnik-et-al-2003,abergel-et-al-2011,juvela-et-al-2012, roy-et-al-2013}.
Additionally, the mid-IR emission, which comes mainly from stochastically heated very small grains \citep[e.g.,][]{draine-li-2001,compiegne-et-al-2011, jones-et-al-2013}, decreases by 80$-$100\% from diffuse to denser regions \citep[e.g.,][]{laureijs-et-al-1991, stepnik-et-al-2003,juvela-et-al-2012}. 
In many of these papers, an anti-correlation of the dust colour temperature $T$ with the opacity spectral index $\beta$ has been observed with increasing column density in the environment \citep{dupac-et-al-2003, stepnik-et-al-2003, desert-et-al-2008, paradis-et-al-2010, veneziani-et-al-2010,bracco-et-al-2011,abergel-et-al-2011, juvela-et-al-2013,sadavoy-et-al-2013}.
Values for $\beta $ and $T$ are derived by fitting observations at wavelengths longer than around 60 $\mu$m with a modified blackbody.
As was shown by \citet{juvela-et-al-2013} the values derived for $\beta$ and $T$ depend strongly on the fitting method and the wavelength range of the observations where the modified blackbody has been fitted.
In general, the $\chi^2$ approach overestimates $\beta$ and underestimates $T$ when fitting noisy observations.
In the literature, we therefore find a wide range for values of $\beta$, from 0.8 \citep{dupac-et-al-2003} to 5 \citep{veneziani-et-al-2010}, and of $T$, from 7 K \citep{desert-et-al-2008, paradis-et-al-2010} to 80 K \citep{dupac-et-al-2003}.
A systematic study of the various fitting methods ($\chi^2$, hierarchical models, and the Bayesian method) was carried out by \citet{juvela-et-al-2013}. Their main result was that the observed anti-correlation cannot be explained only by bias in the fitting methods but must originate, at least partly, in intrinsic variations in dust properties.
It is usually assumed that all four changes in the observed SEDs occur because of the evolution of dust from diffuse to dense regions.
These variations result from the evolutionary processes acting on dust as it responds to its environment.

Changes in the environment include variations in the local gas density and in the radiation field.
Coagulation, accretion, fragmentation, sputtering, and photoevaporation change the properties of dust grains, such as their structure, shape, material composition, and size.
This results in a modification of the optical properties of the grains and therefore to changes in the spectral energy distribution (SED).
Variations in dust optical properties resulting from coagulation have been explored in many papers \citep[e.g.][]{bazell-dwek-1990, ossenkopf-1993, ossenkopf-henning-1994, stognienko-et-al-1995, dwek-1997, fogel-leung-1998,ormel-et-al-2009, ormel-et-al-2011,koehler-et-al-2011,koehler-et-al-2012}.
In the transition from the diffuse ISM to denser regions, the main evolutionary processes are the accretion of mantles composed mainly of amorphous carbon, water ice, and other molecular species \citep{jones-et-al-1990,mathis-1992, whittet-et-al-1996,del-burgo-et-al-2003, kiss-et-al-2006} and grain coagulation \citep{bernard-et-al-1999, del-burgo-et-al-2003, stepnik-et-al-2003, kiss-et-al-2006, ridderstad-et-al-2006, paradis-et-al-2009,ysard-et-al-2013}. Observational evidence of grain coagulation in denser clouds was found by \citet{jura-1980} through an analysis of the extinction towards $\rho$ Oph.

The $\beta$-T anti-correlation is seen for amorphous carbon and silicate materials in laboratory measurements \citep{agladze-et-al-1996,mennella-et-al-1998,boudet-et-al-2005,coupeaud-et-al-2011} and the DCD/TLS model \citep{meny-et-al-2007}.
We do not take variations of $\beta$ with grain temperatures into account in our model, but plan to in future studies. 
Since these variations occur mainly at long wavelengths and for large temperature variations, the influence on the SED and extinction are likely to be small for our assumed dust composition and size distribution.

However, so far, no dust model or dust-evolution model have been able to self-consistently explain all four observed changes in the SED from the diffuse ISM to denser regions, i.e. changes in temperature, spectral index (and their anti-correlation), opacity (emissivity) and the mid-infrared emission. 
Model calculations taking into account the coagulation of separate grains into aggregates could explain the decrease in temperature and the increase in opacity, but were not able to explain the observed increase in $\beta$ \citep[e.g.][]{koehler-et-al-2012}.
This theoretical study was later confirmed by simulations including radiative transfer calculations \citep{ysard-et-al-2012} and the analysis of {\it Herschel} observations of a dense filament in the Taurus molecular cloud complex \citep{ysard-et-al-2013}.
These three studies were based on the \citet{compiegne-et-al-2011} dust model, which consists of neutral and charged PAHs, very small grains (VSGs, $\sim$1.2$-$15 nm in radius)  of amorphous carbon \citep{zubko-et-al-1996} and big grains (BGs, 15$-$110 nm in radius) of astronomical silicate \citep{draine-lee-1984} and of amorphous carbon \citep{zubko-et-al-1996}.

Recently, new optical constants for hydrogenated amorphous carbon were derived \citep{jones-2012a,jones-2012b,jones-2012c}. 
A new dust model that uses these materials, in conjunction with new silicate data, explains the observational data of dust in the diffuse ISM, such as the IR to FUV extinction and the dust SED \citep{jones-et-al-2013,koehler-et-al-2014,ysard-et-al-2015}.
In the present study, we use this new dust model and coagulate the grains into aggregates following the same approach as in \citet{koehler-et-al-2012}.
We also consider the accretion of carbonaceous material and ice from the gas phase onto the surfaces of the aggregates.

This paper is organised as follows: In Sec. \ref{sec:2} we present the dust model and dust evolution. In Sec. \ref{sec:3} we describe the method and calculations and present the results in Sec. \ref{sec:4}. We summarise and conclude our study in Sec. \ref{summary}.


\begin{center}
\begin{figure}[ht]
\begin{center}
\includegraphics[width=0.45\textwidth]{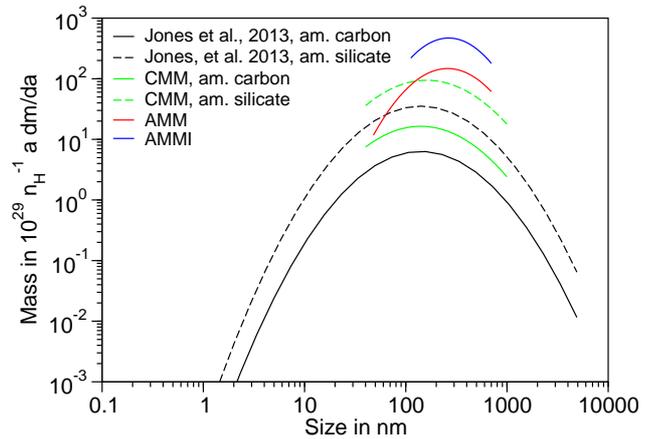}
\caption[]{The mass distribution of the CM dust grains as described in \citet{jones-et-al-2013}, of core-mantle-mantle (CMM) grains and of aggregates (AMM) and aggregates with an accreted ice mantle (AMMI).}
\label{fig:0}
\end{center}
\end{figure}
\end{center}

\section{Dust model and dust evolution}
\label{sec:2}

Our study is based on the recently published dust model from \citet{jones-et-al-2013} for the diffuse ISM ($n_{\rm H}<100~{\rm H/cm^{3}}$).
In the diffuse ISM, we assume that VSGs and BGs are mainly separated.
Very small grains up to about 20 nm in radius consist purely of aromatic-rich amorphous carbon because of UV photo-processing.
Larger VSGs and BGs of amorphous carbon show a core-mantle structure, where the mantle with a thickness of $\sim$20 nm is aromatic rich, while the core is aliphatic rich.
Another BG population consists of amorphous silicates, with Fe/FeS nano-particle inclusions and a mantle of aromatic-rich amorphous carbon, which is formed by accretion and the coagulation of VSGs and has a thickness of $\sim$5 nm.
We assume amorphous silicates of forsterite-type and of enstatite-type and the inclusions occupy a volume of 10\% and consist of a mixture of 30\% FeS and 70\% Fe as described in \citet{koehler-et-al-2014}.
Starting from these grain populations, we consider dust evolution based on the processes of accretion and coagulation, which we assume to occur in relatively dense interstellar environments ($n_{\rm H}>1500$ ${\rm H/cm^{3}}$, as in \citet{ysard-et-al-2013}).

\begin{center}
\begin{figure*}[ht]
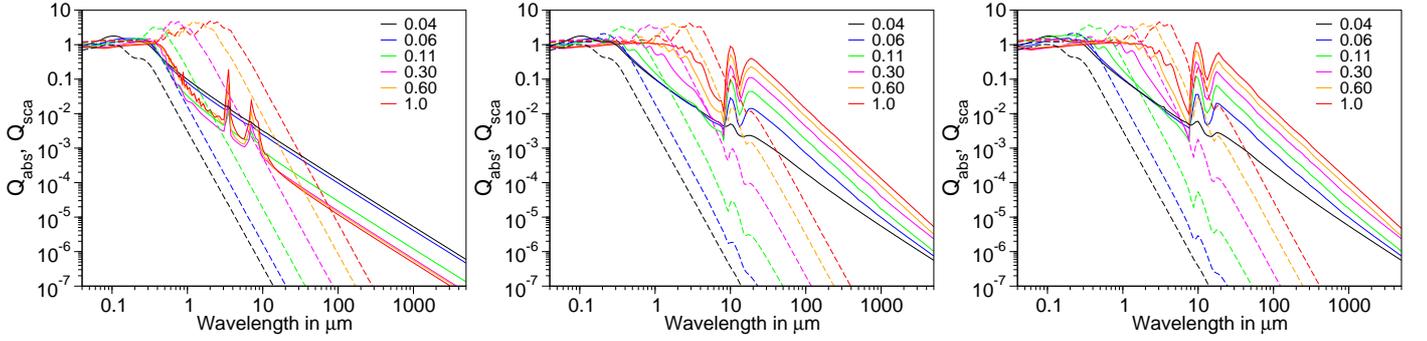

\begin{center}
\includegraphics[width=0.33\textwidth]{Qabs-Qsca-CMM-amCarbon.eps}
\includegraphics[width=0.33\textwidth]{Qabs-Qsca-CMM-amFo10Fe30FeS.eps}
\includegraphics[width=0.33\textwidth]{Qabs-Qsca-CMM-amEnst10Fe30FeS.eps}
\caption{$Q_{\rm abs}$ (solid lines) and $Q_{\rm sca}$ (dashed lines) for CMM grains with cores of amorphous carbon (left) and cores of amorphous silicate of olivine type (middle) and pyroxene type (right) for different grain radii in $\mu$m. The inner mantle is aromatic-rich and the outer mantle aliphatic-rich amorphous carbon. At long wavelengths  $Q_{\rm abs}$ and $Q_{\rm sca}$ increase with increasing grain size.}
\label{fig:1a}
\end{center}
\end{figure*}
\end{center}

\begin{center}
\begin{figure*}[ht]
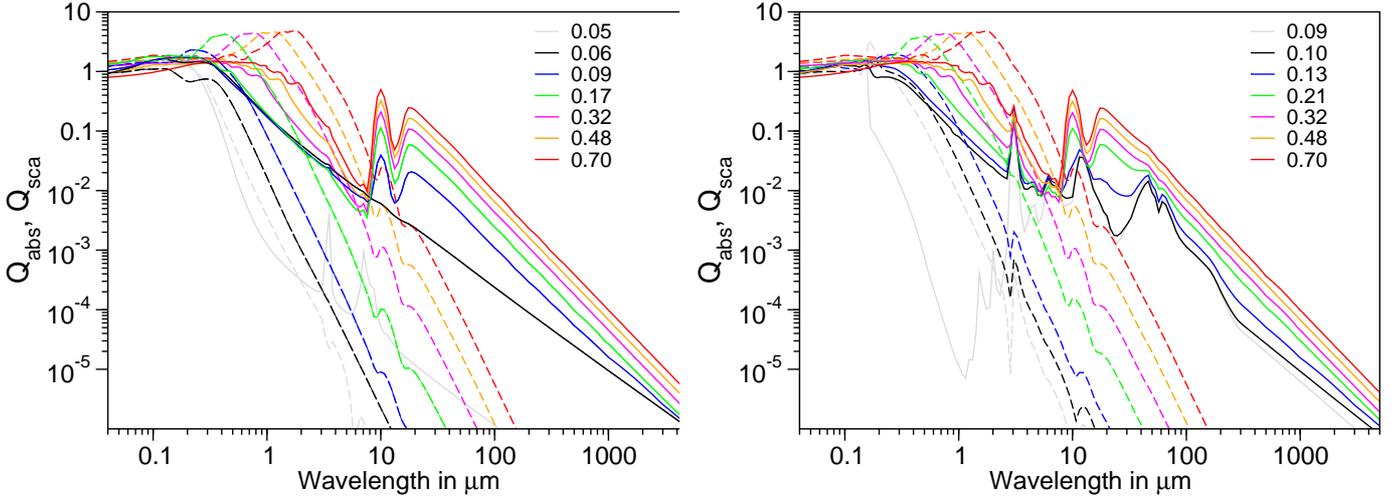

\begin{center}
\includegraphics[width=0.49\textwidth]{Q-AMM-kl.eps}
\includegraphics[width=0.49\textwidth]{Q-AMMI-kl.eps}
\caption[]{$Q_{\rm abs}$ (solid lines) and $Q_{\rm sca}$ (dashed lines) of AMM structure (left) and of AMMI structure (right) for different radii in $\mu$m of volume-equivalent spheres $a_{\rm V}$. At long wavelengths $Q_{\rm abs}$ and $Q_{\rm sca}$ increase with increasing grain size.}
\label{fig:1}
\end{center}
\end{figure*}
\end{center}


\subsection{Accretion of carbonaceous mantles} \label{sec:acc}

The first process that we consider is accretion. Because of the accretion of C and H atoms from the gas phase, an additional amorphous carbon mantle is formed on the surface of grains \citep{jones-et-al-2014}. 
\citet{parvathi-et-al-2012} show that more carbon is available than usually assumed \citep[see e.g.][]{compiegne-et-al-2011}, so that there appears to be sufficient gas phase carbon to form such a carbonaceous mantle.
In low-density regions, this newly accreted mantle would aromatise quickly because of UV photo-processing \citep{jones-et-al-2014}. 
As a result, the grain properties would not change dramatically, since the aromatic amorphous mantle would only increase the particle size but would not change the material properties. 
This would have the strongest influence for the small grains, where the mantle would increase the particle size by a larger factor. 
However, this case was already studied in \citet{ysard-et-al-2015}. 
In denser regions, where UV photons are significantly attenuated (A$_{\rm V}\gtrsim1$), the accretion of C and H atoms will result in an aliphatic-rich amorphous carbon mantle \citep{jones-et-al-2014}, so that BGs will have a core-mantle-mantle structure. 
We investigate how the optical properties of the grains change because of the accretion of such an aliphatic-rich amorphous carbon mantle.

In the following, we use abbreviations to describe the particles with accreted mantles:\\
{\bf CM:} Core-mantle grains from the \citet{jones-et-al-2013} and \citet{koehler-et-al-2014} model.\\
{\bf CMM:} Core-mantle-mantle grains, which are the original grains with an additional aliphatic-rich mantle.

\begin{center}
\begin{figure*}[ht]
\begin{center}
\includegraphics[width=0.99\textwidth]{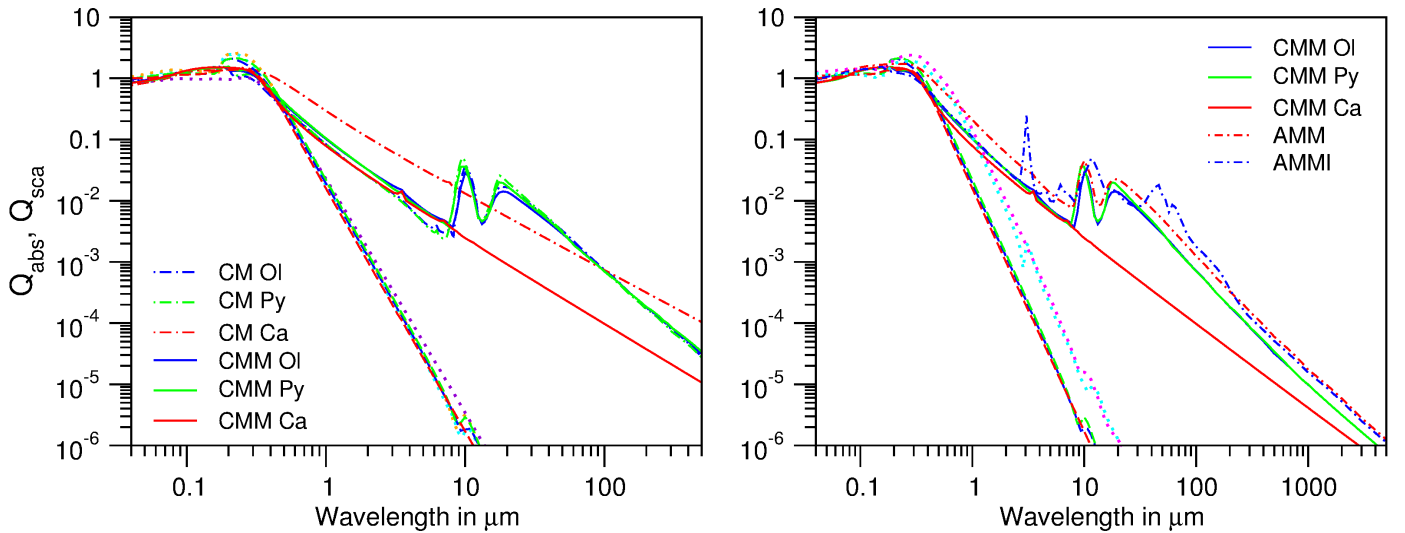}
\caption[]{The differences in $Q_{\rm abs}/{a_V}$ and $Q_{\rm sca}/{a_V}$ for the evolution of grains with CM grain radii of 0.06 $\mu$m. Left: $Q_{\rm abs}/{a_V}$ (dashed-dotted and solid lines) and $Q_{\rm sca}/{a_V}$ (dotted and dashed lines) for CM and CMM grains, respectively. Right: $Q_{\rm abs}/{a_V}$ (solid and dashed-dotted lines) and $Q_{\rm sca}/{a_V}$ (dashed and dotted lines) for CMM grains (as in left pannel) and for AMM and AMMI grains, respectively. Ol indicates amorphous silicate of olivine-type, Py indicates amorphous silicate of pyroxene-type and Ca indicates amorphous carbon.}
\label{fig:4}
\end{center}
\end{figure*}
\end{center}


\subsection{Coagulation into aggregates} 

In denser regions of the ISM ($n_{\rm H}>10^3~{\rm H/cm^{3}}$, A$_{\rm V}\gtrsim1$), grains are expected to coagulate into aggregates. 
Considering the time scales \citep[see e.g.][]{koehler-et-al-2012}, the coagulation of VSGs onto the surface of BGs is relatively fast, so that the mantles on the BGs thicken with time. 
We assume that all VSGs coagulate onto the surfaces of BGs and that the population of VSGs is therefore incorporated into the BG population aggregates.
The time scales to coagulate a few BGs into aggregates \citep[$t=4\times10^5-10^6$ yr,][]{koehler-et-al-2012} are consistent with typical cloud lifetimes \citep[$t\approx10^7$ years,][]{walmsley-et-al-2004}. 
Furthermore, amorphous carbon mantles tend to favour the coagulation of BGs, since they act as a sort of glue \citep{chockshi-et-al-1993}. 
Observations indicate that the relative velocities between the BGs can be small and hence fragmentation processes minimal \citep{ormel-et-al-2009,pagani-et-al-2010}.

We assume aggregates that are formed of 4 BGs (monomers) in an intermediate elongated shape, as described by \citet{koehler-et-al-2012}. 
At long wavelengths, the difference in optical properties compared to separated grains, is largest for the coagulation of a few monomers and remains rather constant for larger aggregates \citep[see e.g.][]{mackowski-2006,koehler-et-al-2011}. 
The time scales to coagulate 4 BGs are realistic in comparison to the cloud lifetime in relatively dense media ($n_{\rm H} = 4 \times 10^3~{\rm cm}^{-3}$) as shown by \citet{koehler-et-al-2012}. 
With the increasing density of the environment, the coagulation rate between BGs increases so that larger aggregates are able to form. 
Larger aggregates have lower temperatures, but the differences in $\beta$ and the FIR opacity are small compared to smaller aggregates  \citep[see e.g.][]{mackowski-2006,koehler-et-al-2011}. We therefore only consider small aggregates consisting of 4 BGs in this study.

The constituents of the aggregates are one BG of aliphatic-rich amorphous carbon and three BGs of amorphous silicate of olivine and pyroxene type with Fe/FeS inclusions as described above.
This distribution agrees with the mass fractions of BGs in the ISM as derived by \citet{jones-et-al-2013}.
We then consider the accretion of a second mantle in denser regions as in Sec. \ref{sec:acc}. 
This second mantle consists of aliphatic-rich amorphous carbon, which is unprocessed because of the high UV extinction.

With increasing density \citep[line of sight $A_{\rm V} > 1.5$,][]{jones-et-al-2014}, the formation of a water-ice mantle occurs. We therefore assume the formation of such ice mantles on the surface of the grains after forming the aggregate.

In the following we use abbreviations to describe the coagulated aggregates:\\
{\bf AMM:} Aggregates consisting of CMM grains.\\
{\bf AMMI:} Aggregates consisting of CMM grains with an additional ice mantle.


\section{Theory and calculations}\label{sec:3}

Model calculations to derive the optical properties of aggregates are carried out with the discrete-dipole approximation \citep[DDA][]{purcell-pennypacker-1973, draine-1988, draine-flatau-2010}. 
We use the DDA code DDSCAT 7.1.0 \citep[for a detailed description see][]{draine-1988, draine-flatau-2010} including the lattice dispersion relations (LDR) \citep{draine-goodman-1993}. 


\subsection{Accretion} \label{sec:acc}

In order to study the accretion process, we consider grain radii from 0.04 to 1.0 $\mu$m. Up to a radius of 0.441 $\mu$m we use DDA to derive the optical properties.
For DDA calculations, the grains are represented by interacting dipoles. In order to obtain the correct size ratios between BGs and VSGs, we assume that BGs of radius (cores) 0.02 $\mu$m are 7 dipoles across, BGs of 0.06 $\mu$m are 17 dipoles across and all larger BGs are 35 dipoles across. 
Since we want to coagulate these grains into aggregates containing 4 BGs, we cannot consider more dipoles because the DDA calculations are then computationally too expensive.
We assume that 1000 dipoles, 3 nm in radius, form the original mantle plus the first coagulated VSGs of aromatic amorphous carbon and that another 1000 dipoles of aliphatic amorphous carbon, 3 nm in radius, form the accreted mantle on each BG. 
For cores of 0.02, 0.06 and 0.11 $\mu$m, where each dipole is 3 nm in size, 1000 dipoles are distributed on the grain surface. 
For larger BGs, the dipole size is larger than 3 nm, since we keep the number of dipoles across the BG constant but increase the radius.
We therefore reduce the number of dipoles in the mantle when increasing the grain size, so that the mantle mass is constant for each BG.
In order to avoid an antenna effect \citep[see e.g.][]{koehler-et-al-2012}, the mantles form a layer so that chains of dipoles protruding from the surface are prevented.
The optical property results are averaged over 125 grain orientations as discussed in \citet{koehler-et-al-2012}, which allows a relatively fast calculation and gives sufficient accuracy. 

For grains with radii larger than 0.441 $\mu$m, we use the Maxwell-Garnett effective medium theory \citep{bohren-huffman-1983} to derive the best mixture of aromatic and aliphatic amorphous carbon in the mantle and amorphous silicate in the core. 
We then use BHMIE \citep{bohren-huffman-1983} to derive the optical properties for the grains\footnote{Using BHCOAT \citep{bohren-huffman-1983} is not possible, since the mantle on the large grains would be too thin to derive reasonable results.}. 
The best mixture was derived by comparing the optical properties for smaller grains, where DDA results are available. 
We find the best mantle mixture of 50\% aromatic-rich and 50\% aliphatic-rich amorphous carbon in volume and a mantle to core volume relation of 0.08\% to 99.2\%.

The mass distribution of CMM grains is shown in Fig. \ref{fig:0}. 
The range of the mass distribution is determined by $a_{\rm min}$, which is the smallest grain size which is possible when adding the accreted mantle and by $a_{\rm max}$, which is the largest grain size we calculate.
We see clearly an increase in the mass as a result of the accretion of material.
For the CMM grains, we use 182 ppm of C for the first mantle and 224 ppm of C for the second mantle. 
The total of 406 ppm is in good agreement with the carbon abundance \citep{compiegne-et-al-2011, parvathi-et-al-2012}.
We derived the average density to be 1.46 g/cm$^3$ for a core of amorphous carbon and and 1.59 g/cm$^3$ for a core of amorphous silicate.


\begin{center}
\begin{figure}[ht]
\begin{center}
\includegraphics[width=0.45\textwidth]{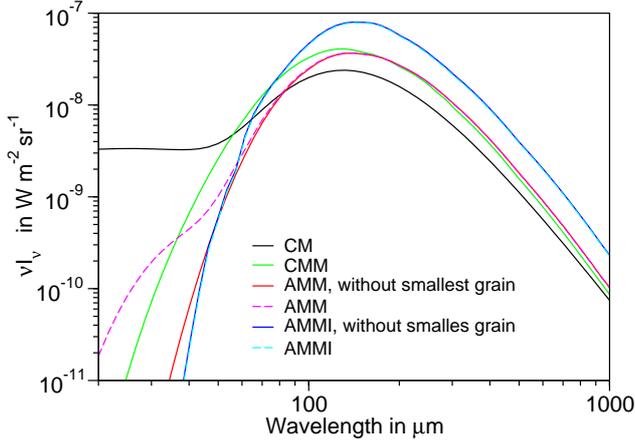}
\caption[]{The SED of the CM grains, CMM grains and aggregates, AMM and AMMI.}
\label{fig:5}
\end{center}
\end{figure}
\end{center}

\subsection{Coagulation}

We now coagulate the CMM grains as described in Sec. \ref{sec:acc} into aggregates of 4 BGs, where the cores of 2 BGs are of amorphous silicate of olivine type, 1 BG of amorphous silicate of pyroxene type and 1 BG of amorphous carbon. 
The optical properties of the aggregates are calculated with DDA for six sizes, with volume-equivalent sphere radii, $a_{\rm V}$, of 0.05, 0.06, 0.09, 0.17, 0.318, 0.476 and 0.70 $\mu$m, averaged over 125 aggregate orientations.
The aggregate size distribution is shown in Fig. \ref{fig:0}.  
The maximum is shifted to larger grains because of the coagulation process.
With our adopted formalism the smallest aggregates, with radius $a_{\rm min}$, actually only consist of carbonaceous mantle materials. 
We assume $a_{\rm max}=0.70$ $\mu$m, since it was computationally too expensive to calculate the optical properties of larger aggregates.
We derive the averaged density to be 1.55 g/cm$^3$.

In a second step, we also study the changes in the optical properties when accreting an ice mantle on the surface of the aggregates. 
For the ice mantles we assume 6000 dipoles of 3 nm in size per BG. We take the optical constants of ice from \citet{warren-1984}.
The mass distribution for these icy aggregates is also shown in Fig. \ref{fig:0}. The $a_{\rm min}$ is again shifted to larger grain sizes. The increase in mass due to the accreted ice mantle is clearly visible.
The average density decreases to 1.18 g/cm$^3$. For the ice mantle we use 500 ppm of O which is in agreement with the abundances given in \citet{compiegne-et-al-2011}.

As shown by \citet{koehler-et-al-2011}, the optical properties of an aggregate depend on the connections between the monomers.
In this study we consider a contact area of 3, 5 and 9 dipoles across for BGs of 7, 17 and 35 dipoles across, respectively.
We also consider that the BGs form the aggregate and VSGs are connected to their surfaces, so that in the contact area the BGs ``touch''. 
As shown by \citet{koehler-et-al-2011}, the increase in the absorption coefficient depends on the real part, $n$, of the optical constants. 
This is at the origin of the increase in opacity (emissivity) at 250 $\mu$m. 
Since the amorphous silicate and the amorphous carbon (E$_{\rm g}$=0.1) have similar values of $n$ at 250 $\mu$m, the differences in the absorption coefficient are small when VSGs provide the contact area.
This was tested by using dimers. 
Initially, we connect silicate BGs and coagulate VSGs of amorphous carbon on their surfaces. 
In follow-up calculations, we separate the silicate BGs by two dipoles and fill the contact area and the surface of the BGs with VSGs of amorphous carbon. 
The number of VSGs is the same in each case. 
The absorption coefficients vary at most by only 4\% and 5\% for aromatic and aliphatic amorphous carbon, respectively. 

\begin{center}
\begin{figure}[ht]
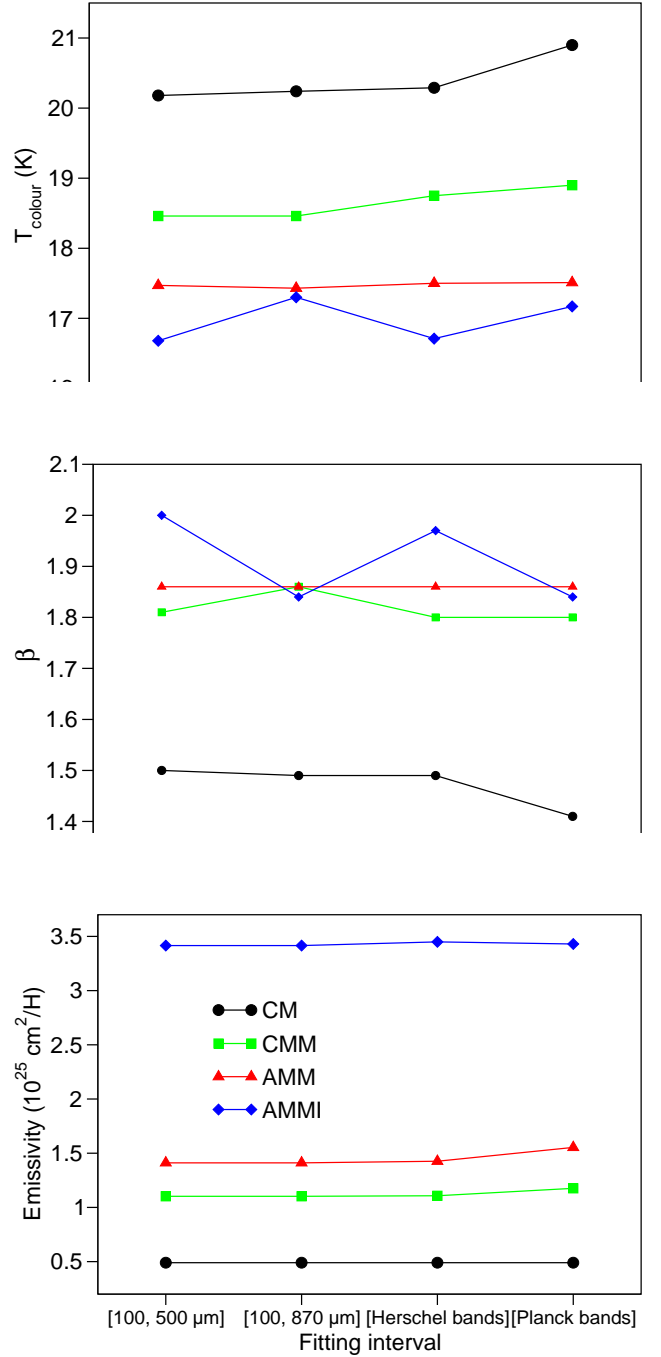

\begin{center}
\includegraphics[width=0.45\textwidth]{Dustem-T-bands.eps}
\includegraphics[width=0.45\textwidth]{Dustem-beta-bands.eps}
\includegraphics[width=0.45\textwidth]{Dustem-emissivity-bands.eps}
\caption[]{Colour temperature, spectral index $\beta$ and opacity derived with DustEM for the \citet{jones-et-al-2013} and \citet{koehler-et-al-2014} model (CM), CMM grains and AMM and AMMI aggregates.}
\label{fig:9}
\end{center}
\end{figure}
\end{center}

\begin{table*}[ht]
\caption{The temperature, spectral index, $\beta$, and increase in opacity derived with the modified blackbody approach, $\epsilon^{\rm BB}_{\rm inc}$, at 250 $\mu$m for the separate grains and aggregates, derived in different wavelength ranges.}
\begin{tabular}{l|ccc|ccc|ccc|ccc}
\hline
\T & \multicolumn{3}{c|}{[100,500]} &  \multicolumn{3}{c|}{[100,870]}  & \multicolumn{3}{c|}{{\it Herschel} PACS \&} & \multicolumn{3}{c}{{\it Planck} HFI \&} \\
 \B &  \multicolumn{3}{c|}{ } &  \multicolumn{3}{c|}{ } & \multicolumn{3}{c|}{IRAS bands} & \multicolumn{3}{c}{SPIRE bands}  \\ 
 \hline
                                   \T \B  & T [K]    & $\beta$  & $\epsilon^{\rm BB}_{\rm inc}$   & T [K]    & $\beta$   & $\epsilon^{\rm BB}_{\rm inc}$ &T [K]    & $\beta$  & $\epsilon^{\rm BB}_{\rm inc}$ &T [K]    & $\beta$ & $\epsilon^{\rm BB}_{\rm inc}$  \\
\hline                                                  
CM                \T      	& 20.2   & 1.5 & 1        	& 20.2   & 1.5  & 1      	& 20.3  & 1.5 &1       	& 20.9  & 1.4 & 1   \\
CMM                         & 18.5   & 1.8 & 2.3     	& 18.5   & 1.9  & 2.3    	& 18.8  & 1.8 &2.3     & 18.9  & 1.8 & 2.4   \\
AMM		   		& 17.5   & 1.9 & 2.9        	& 17.4   & 1.9  &2.9   	& 17.5  & 1.9 & 2.9   	& 17.5  & 1.9  &3.2 \\
AMMI	  \B 		& 16.7   & 2.0 & 7.0        	& 17.3   & 1.8  &7.0   	& 16.7  & 2.0 & 7.0   	& 17.2  & 1.8  &7.0 \\
\hline
\end{tabular}
\label{tab1}
\end{table*}


\subsection{Dust emission}\label{sec:2-2}

Firstly, we use the DustEM code\footnote{DustEM is a publicly available numerical tool to calculate the SED, the extinction and the grain temperature distribution for dust in the ISM, available at http://www.ias.u-psud.fr/DUSTEM/.} \citep{compiegne-et-al-2011} to calculate the SEDs and the extinction of our grains.
We assume a diffuse (transparent) medium illuminated by the interstellar radiation field (ISRF) of \citet{mathis-et-al-1983} for a galacto-centric distance of 10~kpc. 

Secondly, we simulate denser environments by coupling DustEM to the 3D continuum radiative transfer code CRT as described in \citet{juvela-2005} and \citet{ysard-et-al-2012}. 
We consider spherical clouds with the following radial hydrogen density distribution,
\begin{equation}
n_{\rm H}(r) = \frac{n_{\rm C}}{1 + (r/R_{\rm flat})^2} {\rm \;\;\; for \;} R \leq R_{\rm out},
\end{equation}
where $n_{\rm C}$ is the central density, $R_{\rm flat}$ is the internal flat radius, and $R_{\rm out}$ the external radius set to 0.5~pc. 
We define the $(n_{\rm C}, R_{\rm flat})$-parameters for the clouds to be at equilibrium, with their masses per unit length equal to the critical mass defined by \citet{ostriker-1964}, $M_{\rm crit} = 2c_{\rm S}^2/G$, where G is the gravitational constant. 
The sound speed $c_S$ is computed for a gas temperature of 12~K. 
These spherical clouds are then embedded in a more diffuse medium with 100 H/cm$^3$.
We simulate a series of clouds with central visual extinctions of $A_{\rm V} \sim$~0.1 to 20 and illuminate them with the standard ISRF. 

In order to study the variations in dust populations inside the clouds, we consider the following models:\\
{\bf Model CM:} No variation in the dust properties and the diffuse medium as well as the spherical clouds are filled with CM dust of the diffuse-ISM type \citep{jones-et-al-2013,koehler-et-al-2014}.\\
{\bf Model CMM:} The diffuse medium is filled with CM dust while the spherical clouds are filled with CMM dust.\\
{\bf Model AMM:} The diffuse medium is filled with CM dust, and the dense cloud with CMM dust for n$_{\rm H}<$1500 H/cm$^3$ and AMM for higher densities.\\
{\bf Model AMMI:} Same as Model AMM, with the central region filled with AMMI grains.

To compare our dust models with observational results, we finally fit our modelled SEDs with a modified blackbody,
\begin{equation}
S_{\nu} \propto \nu^{\beta} B_{\nu}(T),
\end{equation}
where $B_{\nu}$ is the Planck function, $\beta$ the spectral index, and $T$ the colour temperature. We perform the fits in the {\it Planck}-HFI bands at 857, 545, 353, and 143~GHz, and the IRAS 100~$\mu$m filter. 
We also make fits to the {\it Herschel} PACS and SPIRE filters at 100, 160, 250, 350, and 500~$\mu$m. 
A last pair of fits is performed, without instrumental filters, continuously between 100 and 500 $\mu$m, or 100 and 870 $\mu$m. We do standard weighted least square fits, where all bands/wavelengths are given an equal weight (i.e. we do not consider the effect of noise), in order to determine $T$ and $\beta$.


\section{Results of model calculations}
\label{sec:4}

\subsection{Absorption and scattering coefficients}\label{sec:3-1}

In Fig. \ref{fig:1a} the calculated absorption coefficients $Q_{\rm abs}$ and scattering coefficients $Q_{\rm sca}$ for the CMM grains with different sizes and different core materials (amorphous carbon left, amorphous silicate olivine type middle and pyroxene type right) are presented. 
A difference in $Q_{\rm abs}$ and $Q_{\rm sca}$ with grain size is clearly visible for all core materials. 
The peak in $Q_{\rm sca}$ shifts to longer wavelengths with increasing particle size. 
$Q_{\rm abs}$ increases with increasing grain size for amorphous silicate cores and decreases at wavelengths longer than 1 $\mu$m for the amorphous carbon core.
This decrease is due to the more dominant core material when the grain sizes increase. 
For amorphous carbon cores an increase in the features at 3.4 and around 7 $\mu$m can be seen.
For grains with an amorphous silicate core, an increase in the strength of the silicate features can be observed and differences in the slope at long wavelengths are evident ($\lambda>50$ $\mu$m).

In Fig. \ref{fig:1} $Q_{\rm abs}$ and $Q_{\rm sca}$ are presented for different sized aggregates, AMM (left) and AMMI (right).
With increasing grain size both $Q_{\rm abs}$ and $Q_{\rm sca}$ increase. 
With increasing grain size the silicate features become more pronounced.
The differences in the slope at long wavelengths are strongest for small aggregates.
The smallest aggregates only consists of amorphous carbon material and therefore show strong hydrocarbon features in the spectrum.
For AMMI grains, the ice features are clearly visible for small aggregates and become less evident with increasing grain size while the silicate features become more pronounced.
The smallest grains consist of the pure carbonaceous AMM with a thick ice mantle and we find that the ice features clearly dominate the spectrum.

In Fig. \ref{fig:4}, we present the variation in the optical properties due to the evolution of CM grains with a radius of $a=0.06$ $\mu$m.
It can be seen that $Q_{\rm abs}$ and $Q_{\rm sca}$ vary with accretion (left) and coagulation (right). 
The accretion of an aliphatic amorphous carbon mantle on the amorphous carbon grains shows a decrease in $Q_{\rm abs}$ and a change in slope at long wavelengths, while differences in $Q_{\rm sca}$ are rather small. 
For grains with an amorphous silicate core, we find a decrease in the silicate feature and an increase between 5 and 10 $\mu$m in $Q_{\rm abs}$ due to the accretion of carbonaceous material. 
For AMM grains, $Q_{\rm abs}$ and $Q_{\rm sca}$ increase compared to the single grains, but the overall shape of the curve is comparable to the curves of the amorphous silicate CMM grains. 
For AMMI grains, the ice features are clearly visible in $Q_{\rm abs}$. 
Compared to AMM grains, $Q_{\rm abs}$ decreases between 0.1 and $\sim$5 $\mu$m and for wavelength larger than 200 $\mu$m.
For $Q_{\rm sca}$ the differences are small compared to the AMM grains.

\begin{center}
\begin{figure*}[ht]
\begin{center}
\includegraphics[width=0.99\textwidth]{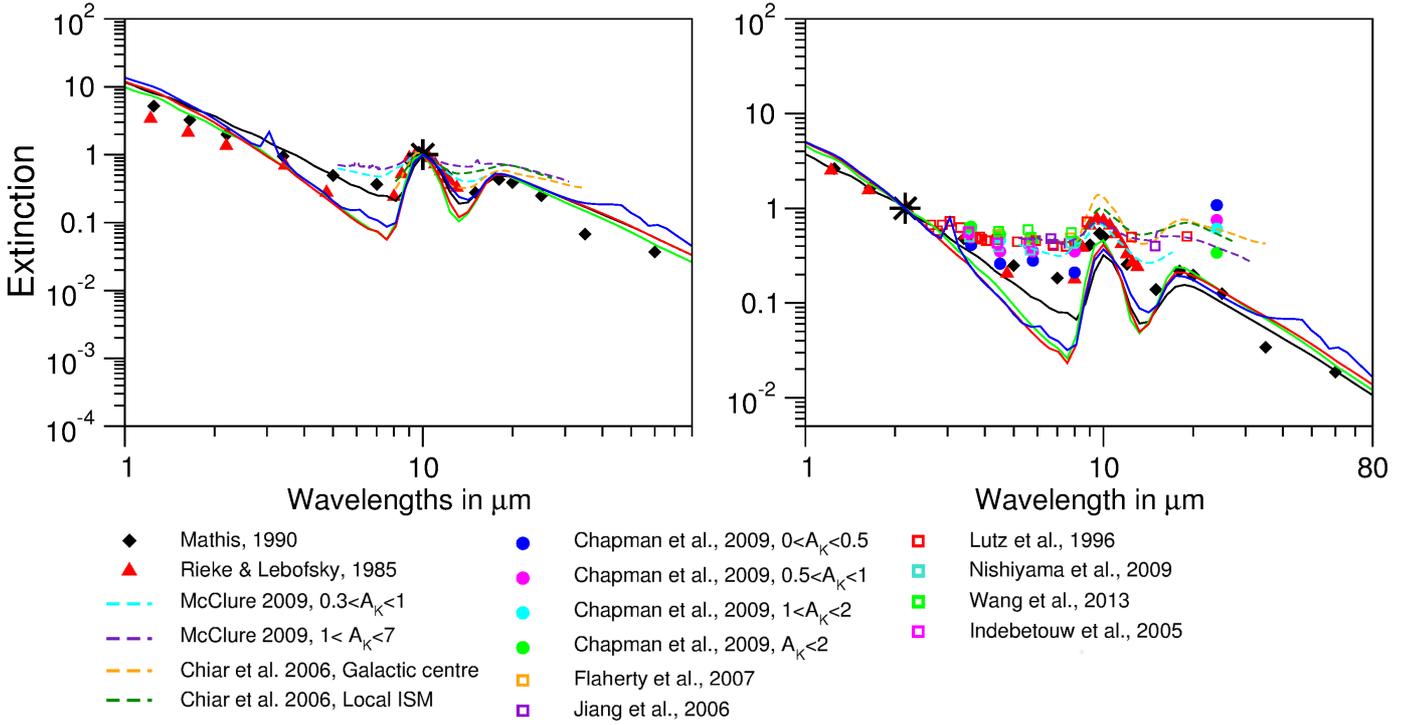}
\caption[]{Infrared extinction of single grains and aggregates normalised at 10 $\mu$m (left) and at the K-band (right) compared to observational data. The normalisation wavelength is indicated by the star. Solid lines: CM grains (black), CMM grains (green), AMM grains (red), AMMI grains (blue).}
\label{fig:7}
\end{center}
\end{figure*}
\end{center}


\subsection{Temperature, spectral index, and opacity derived with modified blackbody fits}\label{sec:3-2}

Using the optical properties described in the previous section, we calculate the corresponding SEDs for grains illuminated by the standard ISRF \citep{mathis-et-al-1983} using DustEM \citep{compiegne-et-al-2011}.
The results are presented in Fig. \ref{fig:5} for CM, CMM, AMM and AMMI grains. 
The SED of the CM grains is the one presented in \citet{koehler-et-al-2014}, where VSGs are still separated and contribute to the emission up to $\sim$70 $\mu$m. 
At longer wavelengths, the SED of CMM grains shows an increase in the intensity of the far-IR emission and a shift to shorter wavelengths indicating an increase in temperature due to stronger absorption at short wavelengths compared to CM grains. 
The steepening of the slope (i.e. an increase in the FIR opacity spectral index) is also clearly visible. 

For AMM grains the SED maximum is shifted to longer wavelengths, which is an indication of the decrease in temperature when coagulating grains.
We can also see a small bump at shorter wavelengths (magenta curve) which is due to the smallest grains which only consist of carbonaceous material. 
The temperature of these grains is around 52 K while all other AMM aggregates are colder than 20 K. 
Because of their high temperatures and small intensity these smallest grains do not contribute to the SED at longer wavelengths.
Excluding these smallest grains, show indeed no variation in the SED at the maximum and longer wavelengths (red line). 
The SED of the AMM grains decreases significantly at 70 $\mu$m as observed by e.g. \citet{stepnik-et-al-2003}.

When adding the ice mantle the far-IR emission increases and the emission peak shifts to longer wavelengths. 
Since an ice band occurs at the wavelength, where the maximum is reached, it is not clear if there is a real shift due to a temperature decrease or if the ice band is dominant. 
The calculated temperatures of AMM and AMMI grains show a small decrease of less than 1.5 K for the larger aggregates. 
The temperature of the smallest purely carbonaceous AMM grain decreases to 14 K when adding the ice mantle. 
Excluding these smallest AMMI grains, we find no variations in the SED (blue line compared to cyan line).

Fitting the calculated SEDs with a modified blackbody, we can derive the dust colour temperature, the spectral index and opacity at 250 $\mu$m. 
In Fig. \ref{fig:9} and Table \ref{tab1}, these values are summarised for the SEDs of the different grain populations, integrated over the {\it Herschel} and {\it Planck} bands and over the 100 to 500 $\mu$m and 100 to 870 $\mu$m wavelength ranges. 
The separate BGs have colour temperatures of around 20 K, in agreement with the findings of \citet{planck-XI-2013}, while the colour temperatures of CMM grains decrease to about 18.7 K. 
The SED consists of three different components, each having a different size distribution. The colour temperature is the value to fit the SED with a modified backbody and not the temperature of the grains.
Compared to CM grains CMM grains with an amorphous carbon core are hotter while CMM grains with an amorphous silicate core have similar temperatures. 
The SED therefore does not show a shift of the peak position, but because of the additional change in the spectral slope, the best modified blackbody fit gives a lower temperature.
For the aggregates, the colour temperature is about 2 to 3 K lower than the CM grains. 
Compared to the calculated temperatures for grains larger than 100 nm, the colour temperatures are on average around 1 K higher.
The spectral index increases from around 1.5 for separate grains to about 1.8 for CMM and AMM grains and up to 2 for AMMI grains.
The opacity at 250 $\mu$m for the original model of \citet{jones-et-al-2013} is found to be $0.5 \times 10^{25}~{\rm cm^2/H}$, which is essentially the same as the observed value of $0.49 \times 10^{25}~{\rm cm^2/H}$ \citep{planck-XI-2013,planck-XVII-2013}. 
We see an opacity increase of around 2.3 for CMM grains and 2.9 for AMM grains, which is in good agreement with the observational results discussed in Sect. 1. 
For AMMI grains, the increase is up to a factor of 7, which is due to the increase in mass when accreting the ice mantle.


\subsection{Extinction}\label{sec:3-3}

Using DustEM, we also derive the extinction of the CMM grains and the aggregates and compare them to observations.
In Fig. \ref{fig:7}, we show the derived extinction compared to observations normalised at 10 $\mu$m (left) and at the K-band (right).
The normalisation at 10 $\mu$m, as suggested by \cite{van-breemen-et-al-2011}, shows the variations at different wavelengths in the extinction curves between AMM and AMMI. 
The ice features are visible and an increase between 6 and 8 $\mu$m and between the two silicate features occurs.
The usual normalisation to the K-band is problematic, since we find differences in the dust extinction in this wavelength range when coagulating grains and/or accretion mantles\footnote{As pointed out by \citet{jones-et-al-2013}, it is likely that the extinction in the B and V bands, as we note for the K band here, varies and so normalisation of the extinction curve at the B and V (and K) bands biases the interpretations.}. 
However, in order to compare the modelled results to observations the data have to be normalised.
The comparison to observations show that the original model by \citet{jones-et-al-2013} and \citet{koehler-et-al-2014} agrees well with the observational data for the diffuse ISM from \citet{mathis-1990}.
The silicate features of the aggregates are in agreement with observations from \citet{rieke-lebofsky-1986} but vary at wavelengths shorter than 5 $\mu$m. 
The silicate features observed by \citet{mcclure-2009} are in general much less pronounced than for the model of the diffuse ISM and the aggregates. 
However, the differences between our model and the observational data might be due to the different methods and assumptions used in the normalisation of the observational data. This topic will be discussed in detail in a following paper.
For our model we derive R$_{\rm V}$ values of $3.5\pm0.2$ for CM grains, $3.9\pm0.2$ for CMM grains, and $4.9\pm0.2$ for AMM and AMMI grains. The relatively large uncertainties are due to the appearance of the ripple and interference structures in the extinction curves, which result from the rather narrow size distributions.

\begin{center}
\begin{figure}[ht]
\begin{center}
\includegraphics[width=0.35\textwidth]{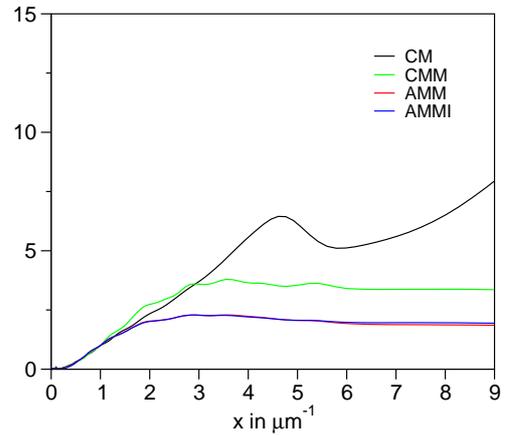}
\caption[]{UV extinction of single grains and aggregates normalised at 1 $\mu$m$^{-1}$.}
\label{fig:8a}
\end{center}
\end{figure}
\end{center}

In Fig. \ref{fig:8a} the extinction normalised to 1 $\mu$m$^{-1}$ in the visible-UV is shown.
For evolved grains, we find strong differences in the UV bump and FUV extinction, which are characteristic of the diffuse ISM and where they appear to arise from small carbonaceous grains (VSGs).
We assume that the VSGs are completely coagulated onto the surface of the BGs and this is why the UV bump disappears for our aggregates.
Coagulation and accretion leads to a strong increase in extinction compared to separate grains. 
We find the same decrease as seen in observations published by \citet{fitzpatrick-massa-2007}, except that in those observations the UV bump is still slightly visible. 
This might be due to VSGs belonging to the diffuse gas surrounding denser regions or distributed along the line of sight and thus contributing to the observed extinction.

\begin{center}
\begin{figure}[ht]
\begin{center}
\includegraphics[width=0.49\textwidth]{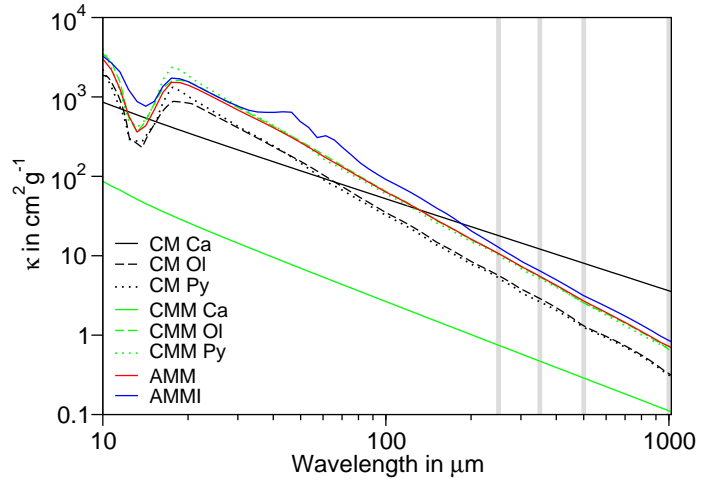}
\caption[]{$\kappa$ shown for the different grains. The grey lines indicate the wavelengths 250, 350 and 500 $\mu$m. Ol indicates amorphous silicate of olivine-type, Py indicates amorphous silicate of pyroxene-type and Ca indicates amorphous carbon.}
\label{fig:8b}
\end{center}
\end{figure}
\end{center}

In Fig. \ref{fig:8b} the mass extinction coefficient $\kappa$ is presented for the original CM grains as well as for the CMM, AMM, and AMMI grains. 
For CM grains the amorphous carbons show the highest $\kappa$ values at the longest wavelengths and the lowest $\kappa$ values with the addition of the aliphatic-rich mantle. For CM grains with amorphous silicate cores, $\kappa$ increases by about a factor of 2 with the addition of the aliphatic-rich mantle. For AMM grains, $\kappa$ is similar to the values for CMM grains with amorphous silicate cores, and increases slightly with the addition of an ice mantle. It is therefore less than ideal to choose a single value of $\kappa$ to derive the total mass of the dust in a galaxy, since dust evolution leads to large differences in $\kappa$. In Table \ref{tab2} we present $\kappa$ for the different models.

\begin{table*}[ht]
\caption{The mass extinction coefficient $\kappa$ in cm$^2/g$ at $\lambda=$ 250, 350 and 500 $\mu$m for the four models presented in the paper. The mass-weighted $\kappa$ is given in the column labelled Tot; it represents the $\kappa$ for extinction and not emission, since the dust temperatures for each component are not taken into account. }
\begin{center}
\begin{tabular}{l|cccc|cccc|c|c}
\hline
\multicolumn{1}{c|}{$\lambda$} \T &  \multicolumn{4}{|c|}{CM } &  \multicolumn{4}{|c|}{CMM}  & AMM & AMMI\\
$\mu$m \B &  Ca &  Ol  & Py & Tot & Ca &  Ol  & Py & Tot &  & \\
\hline                                                  
250      \T &   17.4  & 5.4 & \multicolumn{1}{r|}{5.3} & 8.7 & 0.7 & 10.2 & \multicolumn{1}{r|}{10.2} & 6.4  & 10.4 & 12.5 \\
350         &   11.7 & 2.8 & \multicolumn{1}{r|}{2.6} & 5.3 & 0.5 & 5.1   & \multicolumn{1}{r|}{5.1}     & 3.2  & 5.3   & 6.2 \\
500    \B &     7.8 & 1.3 & \multicolumn{1}{r|}{1.3} & 3.4 & 0.3 & 2.5   & \multicolumn{1}{r|}{2.5}     &  1.5  & 2.6   & 3.1 \\
\hline
\end{tabular}
\end{center}
\label{tab2}
\end{table*}

\begin{center}
\begin{figure*}[ht]
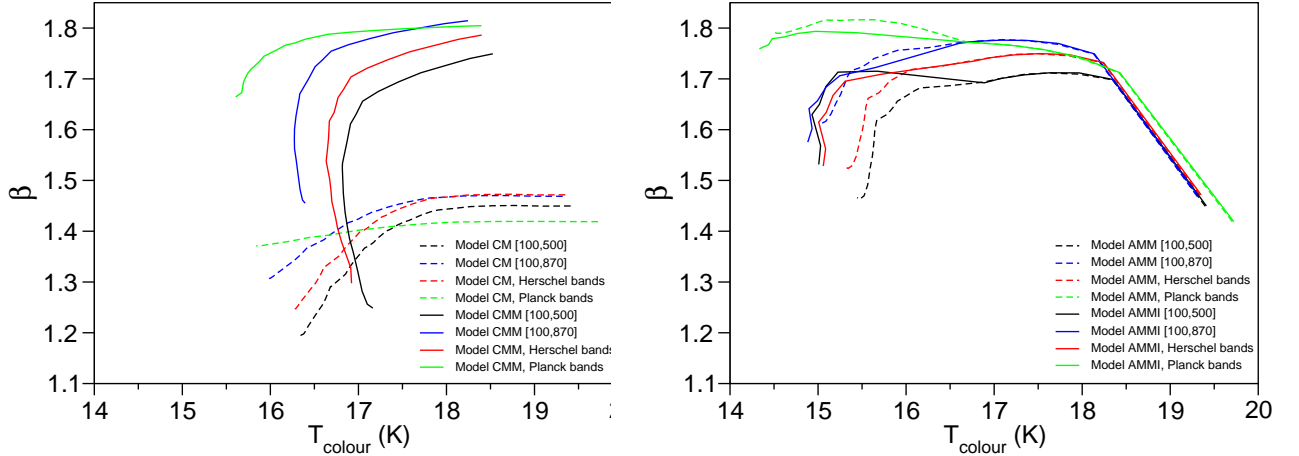

\begin{center}
\includegraphics[width=0.45\textwidth]{RT-beta-T-CM-CMM.eps}
\includegraphics[width=0.45\textwidth]{RT-beta-T-AMM-AMMI.eps}
\caption[]{$\beta$-T correlation for cloud cylinders derived with radiative transfer calculations for the four models described in Section 3.3.}
\label{fig:10}
\end{center}
\end{figure*}
\end{center}


\subsection{Radiative transfer approach}\label{sec:3-4}

We also combine DustEM with the CRT radiative transfer code as described in Sect. 3.3. 
The results for the derived spectral index $\beta$ and colour temperature are presented in Fig. \ref{fig:10} for the four different models presented in Sect. 3.3 and for the analysis with SEDs integrated in the {\it Planck} and {\it Herschel} bands and in the 100 to 500 $\mu$m and 100 to 870 $\mu$m wavelength ranges.

For the CM and CMM grains (Models CM and CMM), $\beta$ decreases with decreasing temperature in all cases. 
This correlation is a pure radiative transfer effect: when the clouds are denser, the mixture of temperatures along the line of sight increases resulting in broader SEDs and thus lower $\beta$ values.
This plot clearly shows the increase in beta, when accreting a second mantle of aliphatic-rich amorphous carbon.
Assuming grain coagulation into aggregates and ice-mantle formation (Models AMM and AMMI), we can clearly see that $\beta$ increases with decreasing temperature down to about 15 K. 
At even lower temperatures, $\beta$ starts to decrease again. 
This means that the radiative transfer effects are strong enough to overcome the increase in $\beta$ of the aggregates.
The $\beta$-T anti-correlation is strong for all considered bands but is more distinct when analysing the data in the {\it Planck} bands, i.e. over a wider wavelength range.
The colour temperatures decrease from around 19.5 to 15 K and $\beta$ increases from 1.45 to 1.8.
These values are reached in regions with $A_{\rm V}\approx16$.
The results match the values measured by \citet{planck-2011} and \citet{planck-XI-2013}.


\section{Conclusions} 
\label{summary}

We investigate how evolutionary processes, i.e. accretion and coagulation, change the optical properties of grains in the transition from the diffuse ISM to denser regions. 
For dust in the diffuse ISM we assume the dust properties of the \citet{jones-et-al-2013} and \citet{koehler-et-al-2014} model.
We also assume that in denser regions these grains undergo evolutionary processes: the accretion of material from the gas phase forms an additional aliphatic-rich mantle of amorphous carbon and the grains then coagulate into aggregates. 
In even denser regions ice mantles form on the surface of the aggregates.
With detailed model calculations we derive the optical properties of these evolved grains and use the DustEM tool and the CRT radiative transfer code to derive the SED and extinction.
With the presented dust-evolution model we are able, for the first time, to self-consistently explain the four observed changes in the SED in going from the diffuse ISM to denser regions with $A_{\rm V}$ of up to 16, namely the changes in temperature, spectral index (and their anti-correlation), opacity (emissivity) and the mid-infrared emission.

The $\beta$-T anti-correlations that we find in the {\it Herschel} bands differ from the observational results. 
The PACS and SPIRE channels have a high level of noise which might significantly influence the measured relation, as shown by \citet{ysard-et-al-2012}. This is not the case for the {\it Planck} HFI+IRAS bands, where we find a good correspondence between the model and observations.
We note that the changes that we observe in $\beta$ are actually driven by gas density-dependent dust evolution processes (accretion and coagulations). Thus, the $\beta$-T effect is actually a $\beta$-$n_{\rm H}$ effect according to the predictions of our model.


\begin{acknowledgements} 
This research acknowledges the support of the French Agence National de la Recherche (ANR) through the program ``CIMMES'' (ANR-11-BS56-0029).
\end{acknowledgements}

\bibliographystyle{aa} 
\bibliography{literatur}

\end{document}